\documentclass[utf8]{FrontiersinHarvard} 

\usepackage{url,hyperref,lineno,microtype,subcaption}
\usepackage[onehalfspacing]{setspace}

\def\keyFont{\fontsize{8}{11}\helveticabold }
\def\firstAuthorLast{Crokidakis et al} 
\def\Authors{Nuno Crokidakis\,$^{1,*}$, Jaime L. C. da C. Filho \,$^{2}$}
\def\Address{$^{1}$Instituto de F\'isica, Universidade Federal Fluminense, Niter\'oi, Rio de Janeiro, Brazil \\
$^{2}$ Instituto Federal do Par{\'a}, Campus Castanhal, Castanhal, Par{\'a},  Brazil
}

\def\corrAuthor{Corresponding Author}

\def\corrEmail{nunocrokidakis@id.uff.br}

\begin{document}
\onecolumn
\firstpage{1}

\title[Corruption as a self-sustained collective state in political systems]{Corruption as a self-sustained collective state in political systems} 


\author[\firstAuthorLast ]{\Authors} 
\address{} 
\correspondance{} 

\extraAuth{}

\maketitle

\begin{abstract}

Political corruption is often interpreted as the result of individual misconduct or isolated institutional failures. However, persistent corruption patterns observed in real-world political systems suggest that systemic corruption may instead emerge as a self-sustaining governance arrangement supported by reinforcing interaction structures. In this work, we introduce a minimal compartmental model for the dynamics of systemic political corruption and the formation of corruption-supporting relational structures. The concentration of political power is treated as an emergent macroscopic observable arising from these coupled dynamics. Despite its simplicity and mean-field character, the model exhibits a nontrivial phase transition separating regimes of low corruption from structurally captured states sustained by self-reinforcing interaction mechanisms. Analytical expressions for the stationary states, the critical threshold and the stability conditions are obtained, revealing how the formation of corruption-supporting structures competes with institutional dissipation mechanisms. The model predicts that, above a critical interaction strength, the captured state becomes dynamically stable, with small perturbations of the macroscopic variables naturally relaxing back to the stationary state. We argue that this mechanism provides a possible explanation for the persistence of corruption across successive electoral cycles and institutional crises. In particular, the political dynamics observed in the state of Rio de Janeiro, Brazil, provide a qualitative illustrative example of several mechanisms discussed by the proposed framework rather than a quantitative application of the model. More broadly, the results suggest that long-term political capture may emerge spontaneously from reinforcing interactions between systemic corruption and the relational structures that sustain it, without requiring centralized coordination or complex strategic behavior.

\section{}


\tiny
\keyFont{ \section{Keywords:} Sociophysics, Corruption dynamics, Collective behavior, Political systems, Nonequilibrium phase transitions, Complex systems}

\end{abstract}

\section{\label{sec:intro}Introduction}

\qquad Corruption has long been recognized as one of the most pervasive and complex social phenomena affecting societies \cite{shleifer1993,jain2001,aidt2003,rose-ackermam2007}. Its impacts extend far beyond economic losses, influencing political stability, institutional trust, social inequality, and the overall functioning of democratic systems \cite{lederman2005,catterberg2006}. Due to its multidimensional nature, corruption has been investigated across several fields of research, including political science \cite{heidenheimer2017}, economics \cite{mauro1997,mo2001} and public administration \cite{monteduro2016}.

Traditional approaches commonly interpret corruption through institutional weaknesses \cite{levitsky2009}, rational choice theory \cite{uhl2025} or governance indicators \cite{donchev2014}. Although these perspectives have provided important insights, they often emphasize individual incentives and institutional structures while overlooking the collective and dynamic nature of corruption processes.

The increasing interconnectedness of contemporary societies has revealed that corruption also exhibits emergent collective characteristics, in which individual decisions are strongly shaped by social interactions \cite{lee2013,villoria2013}. In this context, corruption can be understood not merely as the outcome of isolated strategic decisions, but as a dynamical social process emerging from interactions among many agents \cite{persson2013,rothstein2021}. Such a perspective motivates the application of theoretical frameworks capable of describing collective behavior and non-equilibrium phenomena in complex social systems \cite{persson2013,marquette2015,heydari2024}.

The persistence of corruption and political power has been extensively discussed in political science, particularly in studies of clientelism, coalition formation and political survival \cite{kitschelt2000linkages,mesquita,pappas2009patronage}. These approaches emphasize the role of structured and repeated interactions among political agents, often involving exchanges of support, resources or influence that contribute to the long-term stability of political systems. Clientelistic networks, for example, are characterized by durable relationships that persist across electoral cycles \cite{kitschelt2000linkages,pappas2009patronage}, while coalition theories highlight alliance formation as a central mechanism for maintaining political power \cite{laver1998multiparty}. Similarly, political survival models stress the importance of sustaining stable support structures through the strategic distribution of benefits among key actors \cite{mesquita}. Across these perspectives, a common feature emerges: the persistence of political arrangements is strongly associated with the stability of interaction patterns rather than with the intrinsic attributes of individual agents.

These ideas establish an important conceptual bridge between corruption analyses and statistical physics and sociophysics \cite{castellano2009,parongama2014,serge2011,serge2008,csf_sociophys,sooknanan,serge_frontiers,jusup2022,perc2019,bianconi2023}, fields devoted to understanding how macroscopic patterns emerge from microscopic interactions. Several studies in sociophysics have employed agent-based models \cite{nuno_jorge,vergara2020,elnawawy2022,jaime_nuno_2,conclave,rule_breaking}, network theory \cite{martins2022}, compartmental models \cite{lu2020,birhanu,rahman,tabassum,monteiro,anjam} and nonlinear opinion dynamics approaches \cite{serge_2025,serge_2026} to investigate how collective social behaviors emerge, spread, stabilize or collapse within societies. These approaches suggest that corruption dynamics may exhibit critical behavior and phase transitions.

Despite these advances, the emergence of corruption as a self-sustained dynamical regime driven by feedback between interaction structures and macroscopic social variables remains insufficiently explored. In particular, few studies have formally addressed the existence of critical thresholds separating regimes in which corruption decays from those in which it becomes persistent and self-reinforcing \cite{nuno_jorge,lu2020,birhanu,tabassum}. 

Recent literature on state co-optation argues that organized crime may evolve from an external actor into a structurally embedded component of political institutions \cite{garay}. In such situations, instability is no longer associated exclusively with isolated criminal agents, but may emerge as a collective property sustained by persistent interactions between political, institutional and criminal structures. This motivates the use of approaches inspired by complex systems and statistical physics to study how persistent corruption regimes emerge, stabilize and reorganize through collective interactions.

From this perspective, systemic corruption should not be interpreted merely as a collection of illegal transactions or individual misconduct. Rather, it may constitute a governance arrangement, in which corruption-supporting structures organize access to resources, political influence, institutional protection and coordination among heterogeneous actors. Within this framework, captured political systems reproduce themselves through stable relational structures, providing a natural interpretation for the feedback mechanisms represented in the present model.

In this work, we propose a minimal dynamical framework to investigate the persistence of corruption through a system of coupled ordinary differential equations. The model represents systemic corruption and corruption-supporting relational structures as coupled variables, allowing the description of feedback processes capable of stabilizing captured governance structures over time. These structures are not restricted to formal political alliances; rather, they represent the effective relational infrastructure of mutual support through which corruption becomes structurally embedded and self-sustaining. Their density and persistence determine the macroscopic behavior of the system.

A central contribution of the present approach is the identification of a critical threshold separating regimes of corruption suppression from regimes of self-sustained corruption. The introduction of an effective reproduction number $R_0$ provides a quantitative measure of the balance between reinforcing interactions and dissipative mechanisms, establishing a direct connection between qualitative political science interpretations and quantitative approaches from nonlinear dynamics and statistical physics. From this perspective, the persistence of corruption emerges as a collective dynamical regime arising from interaction structures and feedback mechanisms rather than solely from isolated strategic behavior.


\section{Materials and Methods}

\qquad We propose a minimal nonequilibrium dynamical model to describe the emergence and persistence of systemic corruption in political systems through reinforcing interactions between corruption and support structures among political actors. The system is based on the interplay between two interacting dynamical variables, namely the level of systemic corruption and the formation of corruption-supporting relational structures, and an emergent macroscopic variable describing the concentration of political power.

Let $C(t)$ denote the level of systemic political corruption in the system, interpreted as a macroscopic variable measuring the degree to which corrupt practices are structurally embedded in political and institutional relations. In this sense, $C(t)$ should not be interpreted as a measure of isolated corrupt acts, but rather as an order parameter characterizing the extent of institutional capture sustained by corruption-supporting structures. Let $A(t)$ denote the density of reinforcing support structures among political actors. These structures may include alliance networks, clientelistic ties, patronage relations, protection arrangements or other forms of reciprocal support capable of shielding corruption from institutional dissipation. Depending on the political context, these relational structures may involve both legal and extra-legal forms of coordination among political, bureaucratic, security, economic or criminal actors. Such hybrid governance arrangements have been extensively documented in the literature on criminal governance and plural orders, particularly in urban contexts marked by the coexistence of formal institutions and extra-legal authorities \cite{arias,ariasbarnes2017,lessing2017}. In the present model, these heterogeneous interactions are intentionally coarse-grained into a single macroscopic variable representing the support infrastructure of systemic corruption. Finally, $P(t)$ denotes an emergent macroscopic observable representing the concentration or stability of political power.

The variables are normalized quantities taking values in the interval $[0,1]$. At the microscopic level, $C(t)$ may be associated with the fraction of agents involved in corruption-supporting practices. At the macroscopic level, however, it should be interpreted as an order parameter measuring the degree of institutional embedding of corruption within the political system. Similarly, $A(t)$ can be associated with the density of corruption-supporting relational structures, such as patronage networks, protection arrangements and other forms of reciprocal support among political actors, while $P(t)$ represents an effective measure of their political and institutional influence. In this sense, the model should be interpreted as a coarse-grained description in which the variables summarize the collective outcome of many heterogeneous microscopic interactions.

The time evolution of these variables is governed by the following set of coupled nonlinear differential equations:
\begin{eqnarray} \label{eq1}
\frac{dC(t)}{dt} & = & \alpha A(t)(1 - C(t)) - \beta C(t),  \\ \label{eq2}
\frac{dA(t)}{dt} & = & \gamma C(t)(1 - A(t)) - \delta A(t), \\ \label{eq3}
\frac{dP(t)}{dt} & = & \eta C(t) - \mu P(t). 
\end{eqnarray}

The first equation describes the dynamics of corruption. The term $\alpha A(1-C)$ represents the increase of corruption driven by the presence of support structures, reflecting the idea that relational networks capable of providing protection, coordination and mutual support facilitate the persistence of corruption within political institutions. The factor $(1-C)$ introduces a saturation effect, ensuring that $C$ remains bounded. The term $-\beta C$ accounts for effective institutional dissipation acting on corruption, including legal enforcement, public oversight, institutional accountability and other mechanisms that reduce the persistence of corruption. In this formulation, corruption-supporting alliances act as an endogenous reinforcement mechanism capable of sustaining and amplifying corruption even when corruption levels are initially small. This reflects the idea that stable structures of mutual support and institutional protection may create favorable conditions for the emergence and persistence of corrupt practices.

The second equation governs the evolution of the support structures that sustain corruption. These structures include patronage networks, protection arrangements, clientelistic ties and other forms of reciprocal support that facilitate the persistence of corruption within political institutions. The term $\gamma C(1-A)$ captures the formation of support structures induced by corruption, reflecting the tendency of corruption to generate networks of mutual support that reduce institutional vulnerability and facilitate its persistence. The factor $(1-A)$ again introduces saturation, while $-\delta A$ represents the effective dissipation of corruption-supporting structures, including political competition, institutional disruption, loss of territorial control, fragmentation of support networks, or other mechanisms reducing the persistence of corruption-supporting structures.

The third equation describes the evolution of the emergent macroscopic observable $P(t)$, representing the concentration (or stability) of political power. Unlike $C(t)$ and $A(t)$, $P(t)$ does not participate in the minimal feedback mechanism responsible for the emergence of systemic corruption, but rather quantifies one of its macroscopic consequences. The term $\eta C$ reflects the idea that corruption contributes to the stabilization or concentration of power, for instance through resource accumulation or control over institutional mechanisms. The term $-\mu P$ accounts for natural loss of power due to electoral processes, political turnover or other mechanisms that reduce political influence over time.

Consequently, in the present minimal formulation, the central feedback mechanism under investigation is the mutual reinforcement between systemic corruption and corruption-supporting relational structures. Possible feedback effects from $P(t)$ to $C(t)$ or $A(t)$ are certainly plausible in real political systems, but were intentionally omitted in order to isolate the minimal mechanism responsible for the emergence of self-sustained corruption regimes. Such feedbacks may be incorporated in future extensions of the model.

The parameters $\beta$ and $\delta$ should be interpreted as effective macroscopic dissipation rates. They quantify the net ability of institutional mechanisms to weaken corruption and corruption-supporting structures. In political systems where enforcement agencies are themselves captured or selectively enforced, such effects are naturally incorporated through reduced effective values of $\beta$ and $\delta$, rather than by explicitly modeling each institutional actor.

Corrupt networks frequently seek to consolidate political influence through patronage structures, preferential access to public resources and control over institutional mechanisms. In this sense, corruption may contribute to the accumulation and stabilization of political power over time. Similar mechanisms have been discussed in studies of political survival, path dependence and state capture, where persistent support structures allow political actors to maintain influence beyond individual electoral cycles \cite{garay,mesquita,pierson2000}.

All parameters $\alpha, \beta, \gamma, \delta, \eta$ and $\mu$ are assumed to be positive constants. The parameters $\alpha$, $\beta$, $\gamma$ and $\delta$ characterize the effective reinforcing and dissipative mechanisms governing the coupled corruption-support dynamics, whereas $\eta$ and $\mu$ control the emergence and relaxation of the political power observable. For clarity, Table \ref{tab:parameters} summarizes the interpretation and dynamical role of the effective parameters employed in the model.

\begin{table}[ht]
\caption{Summary of the model parameters and their sociopolitical interpretation.}
\centering
\begin{tabular}{|c|p{5.2cm}|p{5.5cm}|}
\hline
\textbf{Parameter} &
\textbf{Interpretation} &
\textbf{Dynamical role} \\
\hline

$\alpha$ &
Strength of effective corruption reinforcement induced by corruption-supporting relational structures. &
Promotes the increase of systemic corruption. \\
\hline

$\beta$ &
Effective institutional dissipation acting on corruption, including the net effects of legal enforcement, institutional oversight, public scrutiny and other mechanisms that reduce corruption. &
Suppresses systemic corruption. \\
\hline

$\gamma$ &
Strength of effective formation of corruption-supporting relational structures induced by corruption. &
Promotes the growth of corruption-supporting structures. \\
\hline

$\delta$ &
Effective dissipation of corruption-supporting structures due to political competition, institutional disruption, fragmentation of support networks or other destabilizing mechanisms. &
Suppresses corruption-supporting structures. \\
\hline

$\eta$ &
Coupling between systemic corruption and political power concentration. &
Converts corruption into political power. \\
\hline

$\mu$ &
Effective dissipation of political power due to electoral turnover, institutional renewal or loss of political influence. &
Reduces political power concentration. \\
\hline

\end{tabular}
\label{tab:parameters}
\end{table}

It is worth noting that Eq. \eqref{eq3} is linear in $P$ and can be solved explicitly once $C(t)$ is known. In particular, in the stationary regime one obtains
\begin{equation} \label{eq4}
P^* = \frac{\eta}{\mu} C^*,
\end{equation}
\noindent
showing that political power does not introduce an independent dynamical degree of freedom in the stationary regime, being effectively determined by the corruption level. This reflects a deliberate modeling choice in which political power is treated as a macroscopic consequence of corruption and the relational structures supporting it, rather than as an independent driver of corruption dynamics.

Because political power is treated as an emergent macroscopic observable, its transient evolution is determined by the corruption trajectory $C(t)$ together with the initial condition $P(0)$. Therefore, while the stationary political power is uniquely determined by the stationary corruption level through Eq. \eqref{eq4}, different initial values of P may lead to different transient trajectories before converging to the same stationary state.

As a consequence, the essential dynamics of the system is governed by the coupled subsystem defined by Eqs. \eqref{eq1} and \eqref{eq2}. This reduced system captures a feedback loop in which corruption promotes the formation of support structures, such as patronage networks and protection arrangements, and these structures in turn reinforce the persistence of corruption.

Such coupled feedback mechanisms are characteristic of nonequilibrium collective systems exhibiting reinforcing interactions and emergent macroscopic behavior \cite{castellano2009,dickman}. Therefore, the present model focuses on the minimal feedback loop between systemic corruption and the relational infrastructure that sustains it, while political power is incorporated as an emergent variable that quantifies the institutional consequences of the corruption dynamics.

As will be shown below, the model exhibits a transition between a clean phase and a self-sustained corruption regime arising from the competition between reinforcing interactions and effective dissipation mechanisms.

In the following sections, we analyze the stationary states of the system and the conditions under which a persistent, self-sustained corruption regime emerges.


\section{Results}

\qquad We now analyze the stationary solutions and stability properties of the minimal dynamical system defined by Eqs. \eqref{eq1} and \eqref{eq2}. The corresponding stationary behavior of political power then follows directly from Eq. \eqref{eq4}, since $P$ is treated as an emergent macroscopic observable generated by the coupled corruption-support dynamics.


The stationary states are obtained by imposing $dC(t)/dt = 0$ and $dA(t)/dt = 0$ in Eqs. \eqref{eq1} and \eqref{eq2}, leading to
\begin{eqnarray} \label{eq5} 
\alpha A^*(1 - C^*) - \beta C^* & = & 0,  \\ \label{eq6}
\gamma C^*(1 - A^*) - \delta A^* & = & 0. 
\end{eqnarray}
\noindent
where $A^*$ and $C^*$ are the stationary solutions for $A(t)$ and $C(t)$, respectively. From Eq.~(\ref{eq6}), we obtain
\begin{equation} \label{eq7}
A^* = \frac{\gamma C^*}{\delta + \gamma C^*}.
\end{equation}

Substituting Eq.~(\ref{eq7}) into Eq.~(\ref{eq5}), one obtains a nontrivial stationary solution given by
\begin{equation} \label{eq8}
C^* = \frac{\alpha\gamma - \beta\delta}{\gamma(\alpha + \beta)}.
\end{equation}
In addition, the system admits the trivial stationary solution $C^*=0$. Substituting Eq. \eqref{eq8} in \eqref{eq7}, one obtains the nontrivial  stationary solution $A^*$, and the stationary value $P^*$ follows from Eq. \eqref{eq4}, yielding
\begin{equation} \label{eq9}
P^* = \frac{\eta}{\mu} C^*.
\end{equation}

From Eqs. \eqref{eq7} and \eqref{eq9}, the trivial solution $C^*=0$ leads to $A^*=P^*=0$. Thus, two qualitatively distinct stationary states are therefore identified:
\begin{itemize}
\item[(i)] A \emph{clean state}, defined by
\begin{equation} \label{eq10}
C^* = 0, \quad A^* = 0, \quad P^* = 0,
\end{equation}

\item[(ii)] A \emph{captured state}, defined by $C^* > 0$, $A^* > 0$, and $P^* > 0$, which exists only if $\alpha\gamma > \beta\delta$, from Eq. \eqref{eq8}.

\end{itemize}

The clean state should be understood as an idealized analytical reference corresponding to the absence of self-sustained corruption in the model, rather than as an empirically realistic corruption-free political system.

Near the transition, the stationary corruption level behaves as 
\begin{equation} \label{eq12}
C^* \sim \alpha\gamma - \beta\delta .
\end{equation}
\noindent
Therefore, the stationary corruption level $C^*$ plays the role of an order parameter characterizing the transition between clean and captured regimes. In the language of critical phenomena, this scaling relation can be written as $C^*\sim (\alpha-\alpha_c)^{\beta'}$, with critical exponent $\beta'=1$ and the critical corruption reinforcement parameter is given by
\begin{equation} \label{eq13}
\alpha_c = \frac{\beta\delta}{\gamma}
\end{equation}
\noindent
indicating a continuous onset of corruption above the critical threshold. In other words, for $\alpha<\alpha_c$ the valid solution is given by the clean state $(C^*,A^*,P^*)=(0,0,0)$, whereas for $\alpha>\alpha_c$ we have the captured state where $C^*\neq 0, A^*\neq 0$ and $P^*\neq 0$. Thus, the emergence of a persistent corruption regime is governed by a critical threshold above which reinforcing interactions overcome dissipative mechanisms. According to Eq. \eqref{eq13}, the critical threshold results from the competition between effective institutional dissipation acting on corruption ($\beta$) and the ability of systemic corruption to generate and maintain corruption-supporting relational structures, quantified by the parameters $\gamma$ and $\delta$. 

To determine the stability of the stationary states of the minimal dynamical system, we consider the Jacobian matrix associated with the coupled variables $C$ and $A$:
\begin{equation} \label{eq14}
J(C,A) =
\begin{pmatrix}
-\alpha A - \beta &   \,\,\,\,\, \alpha(1-C) \\
\gamma(1-A) &    -\gamma C - \delta
\end{pmatrix}.
\end{equation}

Evaluating the Jacobian at the clean state $(C,A) = (0,0)$, we obtain
\begin{equation} \label{eq15}
J(0,0) =
\begin{pmatrix}
-\beta & \alpha \\
\gamma & -\delta
\end{pmatrix}.
\end{equation}

The corresponding characteristic equation is given by
\begin{equation} \label{eq16}
\lambda^2 + (\beta + \delta)\lambda + (\beta\delta - \alpha\gamma) = 0.
\end{equation}
This equation presents two eigenvalues, given by $\lambda_{\pm}=[(\beta+\delta)/2]\,(-1 \pm \sqrt{1+X})$, where $X=4(\alpha\gamma - \beta\delta)/(\beta+\delta)$. The eigenvalue $\lambda_{-}$ is always negative. On the other hand, $\lambda_{+}$ is negative for $\alpha\gamma < \beta\delta$, i.e., for $\alpha<\beta\delta/\gamma=\alpha_c$, signaling the stability of the clean state. On the other hand, the clean state becomes unstable when $\alpha>\alpha_c$, which coincides with the condition for the existence of the nontrivial stationary solution.

The stability of the captured state can be investigated by evaluating the Jacobian matrix at the nontrivial stationary solution $(C^*,A^*)$, with $C^ *$ and $A^*$ given by Eqs. \eqref{eq8} and \eqref{eq7}, respectively. The Jacobian is given by \eqref{eq14}. Evaluating the trace at the captured state yields
\begin{equation}
tr J=-(\alpha A^*+\beta+\gamma C^*+\delta)<0,
\end{equation}
\noindent
since all model parameters are positive, as well as $C^*$ and $A^*$. Furthermore, using the stationary-state relations, the determinant can be written as
\begin{equation}
\det J=(\alpha A^*+\beta)(\gamma C^*+\delta)
\left(1-\frac{1}{R_0}\right),
\end{equation}
\noindent
where
\begin{equation} \label{eq_Ro}
R_0=\frac{\alpha\gamma}{\beta\delta}.
\end{equation}

Therefore, whenever the captured state exists ($R_0>1$), one has $\det J >0$. According to standard linear stability analysis, a negative trace and a positive determinant imply that the eigenvalues of the Jacobian have negative real parts. Consequently, the captured state is linearly stable and acts as an attractor of the dynamics.

This result indicates that, once the system enters the captured regime, small perturbations decay over time and the dynamics naturally relaxes back to the same macroscopic state. In this sense, the persistence of corruption emerges as a collective property sustained by the reinforcing feedback between systemic corruption and the relational structures that protect and reproduce it.

Beyond the local stability analysis presented above, it is also possible to establish an important global property of the two-dimensional dynamical system governing the variables $C$ and $A$. Let us define the vector field $\vec{F}$ associated with Eqs. \eqref{eq1} and \eqref{eq2} as
\begin{equation} \label{eqF1}
\vec{F} = (\alpha\,A\,(1-C) - \beta\,C,\gamma\,C\,(1-A) - \delta\,A).
\end{equation}

The divergence of this vector field is
\begin{equation}
\nabla .\vec{F} = \frac{\partial{\dot{C}}}{\partial{C}} +  \frac{\partial{\dot{A}}}{\partial{A}} = -(\alpha A+\beta+\gamma C+\delta)
\end{equation}

Because all parameters are assumed to be positive and the physically admissible phase space satisfies $0 \leq C \leq 1$ and $0 \leq A \leq 1$, it follows immediately that $\nabla .\vec{F} < 0$ throughout the entire phase space. Therefore, by the Bendixson-Dulac criterion \cite{perko,strogatz}, no periodic orbit or closed invariant trajectory can exist in the physically admissible region. Furthermore, the square $0 \leq C \leq 1$ and $0 \leq A \leq 1$ is positively invariant under the dynamics, so every trajectory remains bounded for all times. Consequently, the Poincar\'e-Bendixson theorem \cite{perko,strogatz}  implies that every bounded trajectory must asymptotically approach an equilibrium point, since periodic orbits have already been excluded by the Bendixson--Dulac criterion.

Therefore, in the parameter region $R_0>1$, where the captured equilibrium exists and has been shown to be linearly stable, the only possible asymptotic behaviors are the equilibrium points. Since the captured equilibrium is the unique stable equilibrium for $R_0>1$, trajectories starting away from the unstable clean state converge to the captured equilibrium. Hence, no persistent oscillatory dynamics or other nontrivial recurrent motions exist within the positive phase space.

The above condition, Eq. \eqref{eq_Ro}, naturally defines an effective reproduction number, which characterizes the balance between reinforcing interactions and dissipative processes. 

In terms of $R_0$, the transition can be expressed as:
\begin{itemize}
\item $R_0 < 1$: the clean state is stable and corruption decays,
\item $R_0 > 1$: the clean state becomes unstable and the system evolves toward a captured state.
\end{itemize}

Near the critical point, the stationary corruption level scales as
\begin{equation}
C^* \sim (R_0 - 1),
\end{equation}
\noindent
indicating again a continuous transition between clean and captured collective states.

Eq. \eqref{eq_Ro} can be rewritten as $R_0=(\frac{\alpha}{\beta})(\frac{\gamma}{\delta})$. The first factor, $(\frac{\alpha}{\beta})$, measures the balance between corruption reinforcement and effective institutional dissipation. The second factor, $(\frac{\gamma}{\delta})$, quantifies the balance between the formation and the effective dissipation of corruption-supporting relational structures. In other words, systemic corruption can only become self-sustaining when corruption reinforcement and structural support jointly overcome the effective dissipation mechanisms acting on both corruption and corruption-supporting structures.

Summarizing, the results show that the persistence of systemic corruption is governed by the competition between reinforcing interactions, encoded by the parameters $\alpha$ and $\gamma$, and effective dissipation mechanisms, represented by $\beta$ and $\delta$. Persistent captured regimes emerge whenever reinforcement dominates dissipation, allowing corruption and its supporting relational infrastructure to become self-sustaining.

When the reinforcing interactions dominate, i.e., $R_0 > 1$, the system enters a self-sustained regime in which corruption becomes a stable collective state. Conversely, when dissipative mechanisms prevail, the system relaxes to a corruption-free state. This transition provides a simple and general mechanism for the emergence of persistent corruption in interacting political systems.

The analytical results obtained above can be summarized in terms of a phase diagram separating clean and captured regimes, as discussed in the next section.


\section{Discussion}

\begin{figure}[t]
\centering
\includegraphics[width=0.6\textwidth]{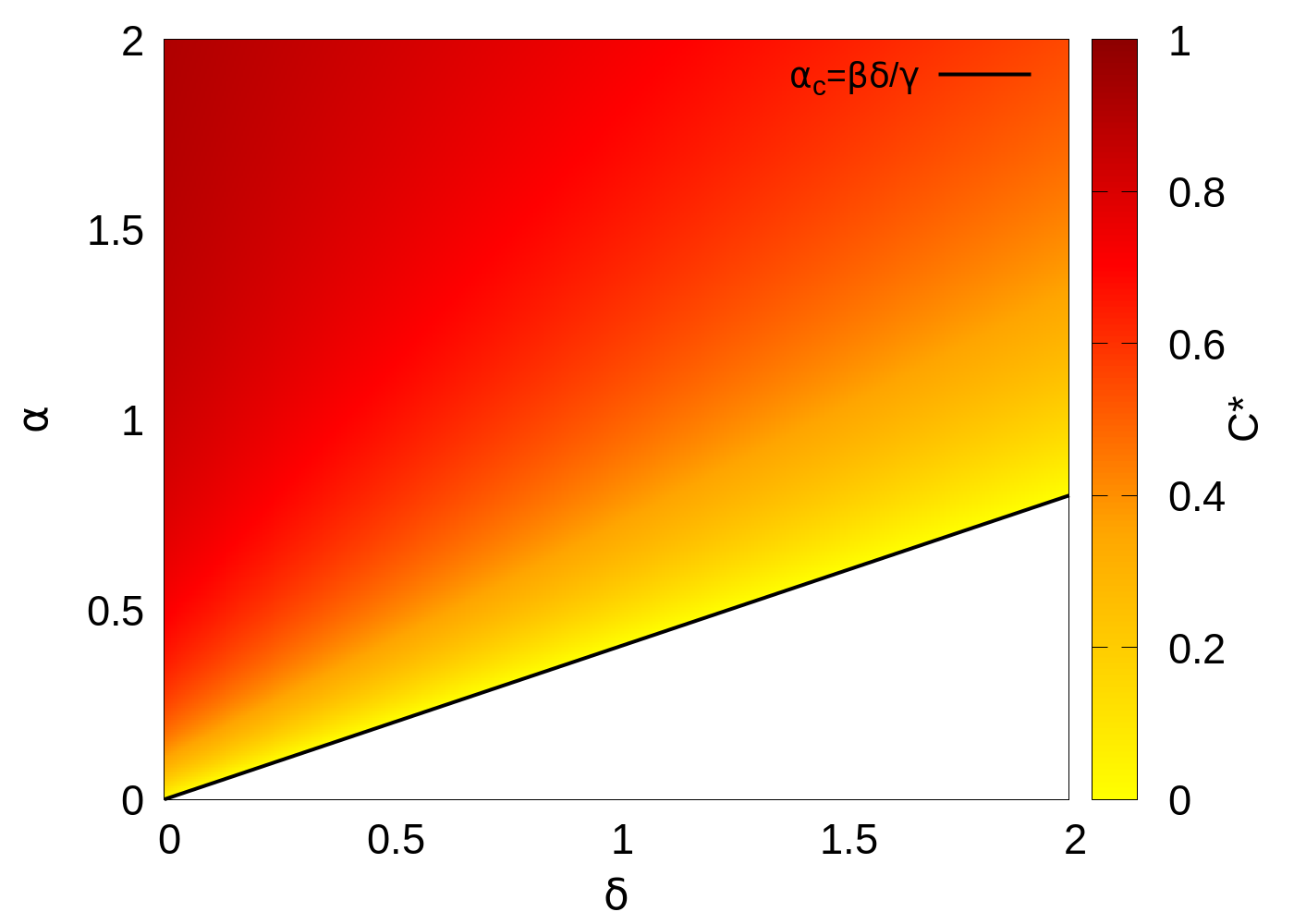}
\caption{Phase diagram in the plane $\alpha$ vs $\delta$. The solid (black) line corresponds to the critical threshold $\alpha_c=\beta\delta/\gamma$, Eq. \eqref{eq13}, separating two distinct dynamical regimes. The white region corresponds to the clean phase, where the stationary corruption level vanishes ($C^*=0$). The colored region corresponds to the captured phase ($C^*>0$), where systemic corruption becomes self-sustained through reinforcing feedbacks between corruption and corruption-supporting relational structures. Color intensity indicates the magnitude of the stationary corruption level, with darker regions representing systems deeper inside the captured phase. The fixed parameters are $\beta=0.2$ and $\gamma=0.5$.}
\label{fig:phase_diagram}
\end{figure}

\qquad The above analytical results naturally define a phase diagram for the system, characterized by the competition between reinforcing interactions and dissipative processes. Such a phase diagram, in the plane $\alpha$ vs $\delta$ is shown in Fig. \ref{fig:phase_diagram}. Here, $\alpha$ quantifies the reinforcing influence of corruption-supporting relational structures on systemic corruption, whereas $\delta$ measures the effective dissipation of these structures. The fixed parameters are $\beta=0.2$ and $\gamma=0.5$. The transition line $\alpha_c=\frac{\beta\delta}{\gamma}$ separates separates a regime in which dissipation dominates from a captured regime sustained by reinforcing interactions.

Below the critical line, dissipative effects dominate and corruption cannot sustain itself in the stationary state, leading to a clean phase (white region in Fig. \ref{fig:phase_diagram}). Above the transition, reinforcing interactions become sufficiently strong to stabilize corruption through collective feedback mechanisms, giving rise to a captured phase (colored region in Fig. \ref{fig:phase_diagram}). The color intensity represents the magnitude of the stationary corruption level, with darker regions corresponding to systems located deeper inside the captured phase.

The use of different color intensities inside the captured phase is motivated by the fact that real-world political systems typically exhibit different corruption levels rather than a strict separation between perfectly clean and fully corrupted states. According to international corruption perception indicators \cite{ti2025}, even countries commonly regarded as highly transparent still display nonzero corruption levels, while others exhibit substantially larger and more persistent corruption patterns. Within the present framework, such differences can be interpreted as systems located at different depths inside the captured phase.

The phase diagram provides a simple but powerful interpretation of political systems in terms of the competition between reinforcement and dissipation. Systems located deep inside the captured phase are expected to remain close to their stationary regime under macroscopic perturbations, whereas systems closer to the critical boundary respond more strongly to institutional changes that modify the underlying interaction mechanisms.

The location in the phase diagram strongly influences the dynamical response of the system. Within the present coarse-grained framework, perturbations are represented as temporary displacements of the macroscopic variables away from their stationary values. Such perturbations may qualitatively correspond to events that transiently reduce the level of systemic corruption, which may qualitatively correspond to institutional interventions, anti-corruption operations or other macroscopic events temporarily reducing the observed corruption level.

For systems near the transition, such perturbations may lead to significant changes in the long-term dynamics. However, for systems deep in the captured phase, the dynamics is dominated by the reinforcing feedback between systemic corruption and the relational structures that protect and reproduce it, causing the system to rapidly relax back to its stationary state.

This behavior provides a natural explanation for the persistence of corruption under repeated institutional interventions or other macroscopic perturbations, as the underlying interaction structure remains largely unaffected.

This interpretation is consistent with empirical observations in systems where corruption persists across multiple electoral cycles, despite repeated interventions \cite{castellano2009,bardhan1997corruption,andvig2001corruption,acemoglu2001colonial,north1990institutions}. In the context of the present model, such behavior is not anomalous, but rather a direct consequence of the system's position in the phase diagram.

Recent political developments may provide illustrative examples of the mechanisms described by the model \cite{misse,arias,pierson2000}. Within the present coarse-grained framework, institutional interventions or anti-corruption operations are interpreted as macroscopic perturbations that temporarily displace the macroscopic state by reducing the observable level of systemic corruption. However, if the underlying corruption-supporting relational structures remain largely preserved, the reinforcing interaction mechanisms responsible for sustaining systemic corruption are not substantially altered. In this case, the system remains within the basin of attraction of the captured state and naturally relaxes back toward the same stationary regime. This interpretation illustrates how captured governance structures may persist despite changes in in political leadership, reflecting the resilience of the underlying support infrastructure rather than the persistence of specific individuals.

The phase diagram highlights that the persistence of corruption is not solely determined by the presence of corrupt actors, but by the balance between structural reinforcement and dissipation. In addition to the mechanisms explicitly represented in the model, real-world political systems may include structural factors that strengthen the emergence and persistence of corruption-supporting relational structures. Such factors effectively enhance the reinforcing interactions represented by the model while weakening the corresponding dissipation mechanisms, thereby favoring the persistence of captured governance structures.

Within the present framework, these effects can be interpreted as an increase in the parameters governing reinforcing interactions and a decrease in those associated with dissipation, ultimately shifting the system deeper into the captured phase.

This perspective emphasizes the importance of structural interventions capable of modifying the reinforcing interaction mechanisms underlying systemic corruption, rather than interventions acting only at the level of isolated political actors or specific institutional episodes.

Empirical observations in some political systems suggest that repeated political turnover and institutional interventions may occur without qualitative modifications of the overall institutional structure \cite{castellano2009,north1990institutions,pierson2000}. Within the present framework, such behavior is qualitatively consistent with the behavior expected for systems located deep within the captured phase, where reinforcing interactions dominate over individual turnover. In this regime, the persistence of corruption is primarily a structural property of the interaction network rather than a consequence of specific individual actors.


\section{Final remarks}

\qquad The present work proposes a minimal dynamical framework for understanding the persistence of corruption as an emergent collective phenomenon. Rather than interpreting corruption exclusively as isolated misconduct, the present framework suggests that persistent corruption may be understood as a self-sustaining governance arrangement emerging from reinforcing interactions between systemic corruption and the relational structures supporting it.

A central result of the model is the existence of a phase transition between qualitatively distinct dynamical regimes. Below the critical threshold, corruption remains limited and dissipative mechanisms dominate the dynamics. Above it, the system evolves toward a structurally stable captured phase sustained by reinforcing feedback interactions between systemic corruption and the support structures that sustain it. In this regime, the model predicts that the macroscopic state becomes comparatively insensitive to macroscopic perturbations, including institutional crises and successive institutional interventions.

More broadly, the phase diagram introduced in this work provides a conceptual tool to compare different political systems in terms of their position relative to the transition. Systems close to the critical boundary may be sensitive to institutional changes, while those far from it require substantial structural modifications to alter their dynamical regime.

These results highlight the importance of the interaction structure underlying political systems. In particular, sustained reductions in corruption require interventions capable of modifying the mechanisms that generate and sustain corruption-supporting structures, rather than focusing exclusively on individual turnover. Otherwise, the system tends to reorganize and recover its previous macroscopic configuration, providing a natural explanation for the persistence of corruption despite repeated enforcement efforts.

Although the present model represents these reinforcing interactions through a single macroscopic variable, the underlying support structures may differ substantially across political systems. In particular, they may involve heterogeneous arrangements connecting politicians, bureaucratic actors, security forces, territorial brokers, economic interests and, in some contexts, organized criminal groups operating through both legal and extra-legal governance mechanisms. The purpose of the present model is not to distinguish these actors explicitly, but rather to capture, in a coarse-grained manner, the collective relational infrastructure capable of sustaining systemic political corruption.

The results presented here highlight a central aspect of complex systems: robust collective structures may emerge from remarkably simple dynamical ingredients. In this sense, the complexity of political systems may arise not from the complexity of individual actors, but from the reinforcing mechanisms connecting them. Within this framework, the persistence of corruption-supporting relational structures can be naturally interpreted as a consequence of structural incentives dominating over purely ideological alignment, allowing captured governance structures to survive individual turnover and institutional interventions.

The political dynamics of the state of Rio de Janeiro, Brazil, provide an illustrative case exhibiting several qualitative features discussed by the present model \cite{misse,pantaleao}. The repeated occurrence of arrests, removals, impeachments and political turnover without evident structural changes in the overall macroscopic organization is qualitatively consistent with the behavior expected for systems operating within the captured regime, where corruption-supporting relational structures remain largely preserved despite changes in political leadership. Within this perspective, coalition reconfigurations often appear driven by strategic access to institutional power and survival incentives rather than by stable ideological identities. Political actors may preserve simultaneous connections with competing groups in order to maintain access to state resources, territorial influence and electoral viability. Such heterogeneous support structures have been extensively discussed in the literature on criminal governance, political co-optation and organized crime penetration into public institutions \cite{garay,arias,pantaleao}. This comparison is intended solely as a qualitative illustration of the proposed theoretical framework. The present work does not attempt to estimate model parameters or provide a quantitative description of the institutional dynamics of Rio de Janeiro.

Although highly simplified, the present approach establishes a bridge between statistical physics, nonlinear dynamics and political science, providing a conceptual framework for the study of persistent institutional structures. Future investigations may incorporate stochastic effects, heterogeneous agents, adaptive interaction networks, or empirical parameter estimation, allowing for a deeper characterization of captured political regimes and their stability properties.

An additional extension would be to allow the effective dissipation parameters to depend on the dynamical state of the system. For example, institutional or structural dissipation may become less effective as corruption-supporting structures consolidate, leading naturally to state-dependent parameters and richer nonlinear dynamics.

\section*{Conflict of Interest Statement}

The authors declare that the research was conducted in the absence of any commercial or financial relationships that could be construed as a potential conflict of interest.

\section*{Author Contributions}

NC: Conceptualization, Data curation, Formal analysis, Funding acquisition, Investigation, Methodology, Project administration, Resources, Software, Supervision, Validation, Visualization, Writing -- original draft, Writing -- review $\&$ editing; JLCCF: Formal analysis, Investigation, Methodology, Validation, Visualization, Writing -- original draft, Writing -- review


\section*{Funding}
Nuno Crokidakis acknowledges partial financial support from the Brazilian scientific funding agency Conselho Nacional de Desenvolvimento Cient{\'i}fico e Tecnol{\'o}gico, Brazil (CNPq, Grants 308643/2023-2 and 406820/2025-2).


\section*{Acknowledgments}
The authors gratefully acknowledge the anonymous reviewers for their insightful comments and suggestions, which substantially improved both the manuscript and its conceptual framework.


\section*{Data Availability Statement}
No datasets were generated or analyzed in this study. The results are based on analytical calculations and numerical evaluation of the derived expressions.


\bibliographystyle{Frontiers-Harvard} 
\bibliography{refs}






\end{document}